\documentclass[ 11pt,a4paper,fleqn]{article}
\usepackage{setspace}
\begin{document}

\title{Flavour physics parameters from  data  and unitarity}

\author{Petre Di\c t\u a\footnote{dita@zeus.theory.nipne.ro}\\
Horia Hulubei National Institute of Physics and Nuclear Engineering\\ P. O. Box MG6, RO-077125, Magurele, Romania}

\maketitle

\begin{doublespace}

\begin{abstract}
The aim of the paper is to show that the nowadays experimental 
data from superallowed  nuclear and neutron  $\beta$ decays, and  from leptonic and semileptonic 
decays allow the  finding of the most probable numerical form of the KM matrix, as well as
 the determination of decay constants, $f_{P}$, and of various form factors  $f_+(q^2)$, by 
using a genuine implementation of unitarity constraints. In particular this approach allows  the 
determination of semileptonic form factors that is illustrated on the existing data
of $D\rightarrow \pi l \nu$ and $D\rightarrow K l\nu$ decays.\\

\end{abstract}

\section{Introduction}
Within the Standard Model (SM) the flavour physics is encoded by the  Kobayashi-Maskawa (KM) matrix,
 \cite{KM}, supposed to be unitary, matrix that  describes the quark flavor mixing through  four
 independent parameters: three mixing angles, $\theta_{ij}$, $ij$ = 12, 13, 23, and one 
\textit{ CP}-violating phase, $\delta$. By consequence the experimental determination of KM 
matrix entries is essential for the validation of the SM, and for  detection of new physics 
beyond it. 

However the determination of KM entries rises a few  problems.
 The first one come from  theory, namely  the mixing angles are not invariant quantities, their 
numerical values depend on the original KM form,   \cite{KM}, or on the present day form,  
\cite{pdg08}, which  is not rephasing invariant,   \cite{HL}. 
 This shortcoming  becomes harmless if one follows Jarlskog's approach. Starting with her first
 papers on  KM matrix,  \cite{J1},  she  proposed the determination of 
 the quark mixing matrix in terms of directly measurable quantities, and, in the same time, 
invariant quantities. In this context an invariant quantity is one whose numerical value does
 not depend on the KM matrix form, or of its rephaising invariance.   Jarlskog  provided two
 such invariants:  the KM moduli, and the celebrated  $J$ invariant, \cite{J2}.  After few years it was realized that all the measurable quantities of 
the quark mixing matrix are expressible in terms of four independent KM matrix moduli,  \cite{BL} and
 \cite{J3}. In these papers it was also shown that  areas of all unitarity triangles are equivalent, 
and  numerically equal to half of the $J$ invariant.
As a conclusion we can state that the numerical values for all the measured invariant quantities  should be 
the same irrespective of the physical processes  where they are involved.

The second difficulty comes from the experimental side. If one uses the KM moduli set as independent 
parameters, these ones are not directly measured by experimenters. In the simplest case, that of leptonic
 decays, they measure branching ratios and provide numbers for  products of the form $|U_{qq'}|f_P$, 
where $U_{qq'}$ is the corresponding KM matrix element, and $f_P$ is the decay constant. For meson 
semileptonic decays the physical observable is the differential decay rate, $d\Gamma/dq^2$, which up to 
known factors, is proportional to  $|U_{qq'}f_+(q^2)|^2$, where  $q$ 
denotes the transferred momentum between initial and final  mesons. In the last case the experimental teams
 usually provide numerical values for  products of the form $|U_{qq'}f_+(0)|$, where $f_+(0)$ is the
 semileptonic decay form factor at zero-momentum transfer. It is clear that from such  measurements one 
cannot find two unknown parameters, say  $|U_{qq'}|$ and $f_P$. This can be done if  and only if one can find independent
 constraints on KM matrix moduli; fortunately  these ones are provided by unitarity.
The main aim of the paper is to show how  the  unitarity property of the KM matrix can be transformed 
into a powerful tool for the determination  of both matrix moduli and form factors directly 
from the experimental data.

The unitarity constraints are presented in Sec. II  where they are implemented in a $\chi^2-$form that depends only on KM matrix moduli. In Sec. III we present the decay formulae for superallowed $0^+\rightarrow 0^+$ nuclear and neutron  $\beta$ decays, and those for leptonic and semileptonic decays. They depend on KM matrix  moduli and  specific decay parameters such as decay constants, $f_P$, and form factors $f(q^2)$, etc, that are implemented in an other $\chi^2$-piece. In Sec. IV we cite the experimental papers from which  we took the data  that we used in our fit, and in Sec. V we present the  numerical results. The paper ends by Conclusion.

\section{Unitarity constraints}

The use of $|U_{ij}|$ as independent parameters raises an important problem, namely the solving of 
 the consistency problem between moduli and unitarity property,
 which all amounts to obtaining the necessary and
 sufficient conditions on  the set of numbers $|U_{ij}|$ 
 to represent  the moduli  of an exact unitary matrix. After that we have to find a device for applying these conditions   to the experimental situation where the data are known modulo uncertainties.

Both these problems have been solved, and a procedure for recovering   KM matrix elements from error 
affected data was provided in \cite{PD}. These unitarity constraints say that the four independent parameters
  $s_{ij}=\sin\theta_{ij}$ and $\cos\delta$  should take physical values, i.e. $s_{ij}\in (0,1)$ 
and $\cos\delta \in (-1,1)$, when they are computed via   equation set: 
\begin{eqnarray}
V_{ud}^2&=&c^2_{12} c^2_{13},\,\, V_{us}^2=s^2_{12}c^2_{13},\,\,V_{ub}^2=s^2_{13}\nonumber \\
 V_{cb}^2&=&s^2_{23} c^2_{13},\,\,
 V_{tb}^2=c^2_{13} c^2_{23},\nonumber\\
V_{cd}^2&=&s^2_{12} c^2_{23}+s^2_{13} s^2_{23} c^2_{12}+2 s_{12}s_{13}s_{23}c_{12}c_{23}\cos\delta,\nonumber\\
V_{cs}^2&=&c^2_{12} c^2_{23}+s^2_{12} s^2_{13} s^2_{23}-2 s_{12}s_{13}s_{23}c_{12}c_{23}\cos\delta,~~~~~\label{uni}\\
V_{td}^2&=&s^2_{13}c^2_{12}c^2_{23}+s^2_{12}s^2_{23}-2 s_{12}s_{13}s_{23}c_{12}c_{23}\cos\delta\nonumber,\\
V_{ts}^2&=&s^2_{12} s^2_{13} c^2_{23}+c^2_{12}s^2_{23} +2 s_{12}s_{13}s_{23}c_{12}c_{23}\cos\delta\nonumber
\end{eqnarray}
The above relations have been  obtained by using the standard KM matrix form, \cite{pdg08},
 where $V_{ij}=|U_{ij}|$, and $U_{ij}$ are the  KM matrix  entries. In the paper \cite{PD} it was  also shown  that if
 the independent parameters are the  KM matrix moduli the reconstruction of a unitary matrix  knowing its moduli 
 is essentially unique. By consequence in the following the used independent parameters in  our
 phenomenological analysis  will be the $V_{ij}$ moduli. Although only four of them can be independent, the 
experimental data that are known only modulo uncertainties ``force'' us to use all the possible sets of four independent moduli, as it will be shown
 in the following. A simple combinatorial evaluation shows that there are 57 such sets.

The relations (\ref{uni}) are rephaising invariant, i.e. they have the same form after    multiplication of  all the 
 KM matrix rows and columns by arbitrary phases. More important is that  they   contain a part of the unitarity 
constraints.  It is easily seen that all the six relations such as 
\begin{eqnarray}
V_{ud}^2+V_{us}^2+V_{ub}^2=1\label{ds}
\end{eqnarray}
are a consequence of the relations (\ref{uni}). The  relations (\ref{ds}) are the necessary conditions for  unitarity 
fulfilment, but they are not \textit{sufficient}, as we show in the following. In fact the relations  (\ref{ds}) tell us  that
 if they are satisfied then there exists a physical solution for the mixing angles $s_{ij}$. The relations (\ref{ds}) are not sufficient because the unitary matrices are naturally embedded in a larger class of matrices: the set of double stochastic matrices, \cite{MO}.

A $3\times 3$ matrix $m$ is said to be  double stochastic if its entries 
satisfy the relations
\begin{eqnarray}
m_{ij}\ge 0,\;\;\sum_{i=1}^3 m_{ij} =1,\;\;\sum_{j=1}^3 m_{ij} =1\end{eqnarray} \\
This set is a convex set, i.e. if $M_1$ and  $M_2$ are double stochastic, then
\begin{eqnarray} M=x M_1+(1-x)M_2,~~~0\le x\le 1\label{con}\end{eqnarray}
is  double stochastic.
 The unitary matrices are a subset of this larger set if we define the $m_{ij}$ entries by the relation $m_{ij}=|U_{ij}|^2$, i.e. one get  relations similar to relation  (\ref{ds}).

For example if we choose four independent  moduli, e.a:
 $V_{us}=a,\,V_{ub}=b,\,\,V_{cb}=c,{\rm and}\, V_{cd}=d$, then the following matrix
\begin{eqnarray}
DS_1=\left[\begin{array}{ccc}
1-a^2-b^2&a^2&b^2\\
d^2&1-c^2-d^2&c^2\\
a^2+b^2-d^2&c^2+d^2-a^2&1-b^2-c^2\label{ds1}\end{array}\right]\end{eqnarray}
is a double stochastic  matrix since it exactly satisfies  all the six relations similar to  relation (\ref{ds}), as it is easily checked.

From any four independent moduli entering  (\ref{uni}) one can get the mixing parameters  $s_{ij}$ and
 $\cos\delta$, i.e. the four independent  parameters entering the KM unitary matrix.
 For example with the above choice we find
 from the first six equations  (\ref{uni})  all the three mixing  parameters, $s_{ij}$, and  $\delta$ as follows
{\small\begin{eqnarray}
 s_{13}&=&V_{ub}=b,\,\,s_{12}=\frac{a}{\sqrt{1-b^2}}, \,\,s_{23}=\frac{c}
{\sqrt{1-b^2}}\label{col}\\
 \cos\delta& =&\frac{(1-b^2)(d^2(1-b^2)-a^2)+c^2(a^2+b^2(a^2+b^2-1))}{2 a b c \sqrt{1-a^2-b^2}\sqrt{1-b^2-c^2}}\nonumber
\end{eqnarray}}\noindent
and from the remaining relations three new  $\cos\delta$ formulae similar to (\ref{col}). 
The above relation shows that $\cos\delta$ is an other invariant in the Jarlskog sense depending on four 
independent moduli, and the {\em CP}-violation phase can be measured via relations such as (\ref{col}).

If we make use of the last four relations (\ref{uni}) we get only one solution for mixing parameters and
 $\cos\delta$. Thus  depending on the chosen four independent moduli set the number of solutions varies between
 one and four. 
Because there are 57 such  groups one get 165 different expressions for $\cos\delta$. They  take 
 the same numerical value when are  computed via Eqs. (\ref{uni}), {\em if and only if all the six relations 
similar to Eq.(\ref{ds}) are exactly satisfied}. If the moduli matrix generated by four independent moduli 
is compatible with unitarity then  $\cos\delta\in(-1,1)$, and outside this  interval when the corresponding 
matrix is not compatible.

To see that we make use of the numerical example
$V_{us}=2257/10^4,\;V_{ub}=359/10^5,\;V_{cd}=2256/10^4,\; {\rm and}\; V_{cb}=415/10^4$. By using the $DS_1$ matrix, (\ref{ds1}), one get the $|U|$ moduli as
\begin{eqnarray}|U|=\left[\begin{array}{ccc}
\frac{\sqrt{9490466219}}{10^5}&\frac{2257}{10^4}&\frac{359}{10^5}\\*[2mm]
\frac{141}{625}&\frac{3\sqrt{10526471}}{10^4}&\frac{83}{2\times 10^3}\\*[2mm]
\frac{\sqrt{580181}}{10^5}&\frac{\sqrt{10982}}{2500}&\frac{\sqrt{9982648619}}{10^5}
\end{array}\right]\label{pdg}\end{eqnarray}
By using the formula from the second row of  (\ref{col}) one gets $\cos\delta \approx 0.64088$, showing that the above moduli matrix, (\ref{pdg}),
 comes from  an exact unitary matrix.

If we modify the previous numerical $V_{us}$ value  by adding to it the small quantity $3\times 10^{-4}$ the
 mixing parameters are still physical, only $s_{12}$ is modified by a very small quantity, and respectively 
 three square root entries from the first two columns of (\ref{pdg}), necessary for the  exact fulfilment  of all the six relations similar to (\ref{ds}).
  In this case one gets
$\cos\delta\approx-1.42427$, which shows that the new  moduli matrix, $|U|$,  is not compatible with unitarity, even
 if all of its entries are positive and  satisfy \textit{exactly} all the six relations (\ref{ds}). 
  If one computes the $J$ invariant one finds in the  above two cases
 \begin{eqnarray}J^2=6.317 \times10^{-10},\;{\rm and}\;  J^2=-1.106 \times10^{-9}\end{eqnarray}

Thus the physical conditions for unitarity compatibility are $\cos\delta\in (-1, 1)$, and $J^2 >  0$, 
 respectively, and from a theoretical point of view they are equivalent.

There are even  bad cases; for example if instead of $a$ and $b$ we choose the parameters $V_{ud}=f,\; {\rm and}\;V_{ts}=g $ the corresponding double stochastic matrix has the form
{\small\begin{eqnarray}
DS_2=\left[\begin{array}{ccc}
f^2&c^2+d^2-g^2&1-c^2-d^2-f^2+g^2\\
d^2&1-c^2-d^2&c^2\\
1-d^2-f^2&g^2&d^2+f^2-g^2\nonumber\end{array}\right]\end{eqnarray}}
With the same numerical values for $V_{cd},\;{\rm and}\; V_{cb}$, and  $V_{ud}=f=\frac{97419}{10^5},\; {\rm and}\;V_{ts}=g=\frac{405}{10^4} $ we find
\begin{eqnarray}
V_{ub}^2= 1-c^2-d^2-f^2+g^2=-\frac{72761}{10^{10}}\end{eqnarray}
and all the other $DS_2$ entries are positive. Thus $V_{ub}$ takes an imaginary value, and  $\cos\delta$ and $J$ take complex   values.

Similar results can be easily obtained; another example is that generated by the moduli, $V_{cd}=d, V_{cs}=h,V_{td}=k,V_{ts}=g$,
 whose corresponding double stochastic matrix has the form
{\small\begin{eqnarray}
DS_3=\left[\begin{array}{ccc}
1-d^2-k^2&1-g^2-h^2&-1+d^2+g^2+h^2+k^2\\
d^2&h^2&1-d^2-h^2\\
k^2&g^2&1-g^2-k^2\nonumber\end{array}\right]\end{eqnarray}}
If we choose $V_{cd}=2252/10^4, V_{cs}=97345/10^5, V_{td}=862/10^5, V_{ts}=403/10^4$ one gets $\cos\delta\approx 0.3685$ which means that the corresponding $|U|$ is compatible with unitarity. If we subtract from  $V_{cs}$ value the tiny quantity $5\times 10^{-6}$ one gets  $\cos\delta\approx -5.0234$, and if we subtract $9.3\times 10^{-6}$ we get $\cos\delta\approx -1.641\,i$.

 For numerical computations the use
 of $\cos\delta$ formulae, like (\ref{col}), seems to be more efficient because of their great  sensitivity 
to small moduli variation, as the above numerical computations show.

The real physical cases are those where the central value moduli matrices, directly determined from data,
 or from a fit do not  exactly satisfy   all the six relations (\ref{ds}), but only approximately; for example for a good
 fit the difference from unity could be of the order, $10^{-5}-10^{-7}$, i.e. rather small from a phenomenological point of view.
 In these cases the different formulae for  $\cos\delta$ provide different values, physical and unphysical, 
 even if the mixing parameters take physical values as in the previous examples. Thus the physical reality obliges
 us to implement the unitarity  constraints
\begin{eqnarray}
 \cos\delta_i\approx\cos\delta_j,\;i\ne j,\; {\rm all}\; \cos\delta_i\in(-1,1) 
\end{eqnarray} 
into a $\chi^2-$fitting device, and our choice is 
 \begin{eqnarray}
\chi^2_{1}=
\sum_{j=u,c,t}\left(
\sum_{i=d,s,b}V_{ji}^2-1\right)^2
+\sum_{j=d,s,b}\left(
\sum_{i=u,c,t}V_{ij}^2-1\right)^2 \nonumber\\
+ \sum_{i < j}(\cos\delta^{(i)} -\cos\delta^{(j)})^2,\,\,\,\,-1\le\cos\delta^{(i)}\le 1~~~~ \label{chi1}
\end{eqnarray}
 that enforces all the  unitarity constraints.

\section{Decay Formalism}

In this section we show how  the used  decay formalism for the description of available experimental
 data  allows us to define a second piece of the $\chi^2$-function by taking into  account as
 much as possible the experimental information.

For example the $V_{ud}$ parameter enters  the description of   superallowed $0^+\rightarrow 0^+$ nuclear $\beta$
 decay, and neutron $\beta$ decay. Superallowed $0^+\rightarrow 0^+$  $\beta$ decay between $T = 1$ analog 
states depends uniquely on the vector part of the weak interaction and, according to the conserved vector 
current  hypothesis, its 
 experimental $f t$ value is related to the vector coupling constant, which 
is a fundamental constant and by consequence has the same value for all such transitions, see \cite{HT1},
 \cite{HT2}, \cite{HT3} and \cite{GS}. This means that the following relation should hold
\begin{eqnarray}
ft=\frac{K}{2|G_V|^2\,|M_F|^2}= \textrm{const}\label{g1}
\end{eqnarray}
where $K/(\hbar c)^6=2\pi^3\,\hbar\,{\rm ln 2}/(m_e c^2)^5$, $G_V$ is
the vector coupling constant for semi-leptonic weak interactions, and
$M_F$ is the Fermi matrix element which in this  case is equal to $\sqrt{2}$.
The 
$ft$ value that characterizes any $\beta$ transition depends on the total transition energy $Q_{EC}$, 
 the  half-life, $t_{1/2}$,  of the parent state, and the branching ratio for the particular studied 
transition, \cite{HT2}. However the above relation is only approximately satisfied by a restricted data set,
 and for this set  one defines a ``corrected'' value 
$\mathcal{F}t\equiv ft(1+\delta^{'}_{R})(1+\delta_{NS}-\delta_C) $, where $\delta^{'}_{R} $ and  $\delta_{NS} $ comprise the transition-dependent part of the radiative correction, while  $\delta_C$ depends on  details of the nuclear structure. In the above formula one takes $|G_V^2|=g_V^2 V_{ud}^2$, with  $g_V =1$, and (\ref{g1}) is  written  in the new form
\begin{eqnarray}
\mathcal{F} t =\frac{K}{2  V_{ud}^2(1+\Delta_R^V)}\label{g3}\end{eqnarray}
where  $\Delta_R^V $ is the  transition-independent part of the radiative corrections whose last estimation 
given in \cite{HT2} is 
\begin{eqnarray}\Delta_R^V =(2.361\pm0.038)\% \label{rad}\end{eqnarray}

Similarly for neutron $\beta$ decay we make use of the formula
\begin{eqnarray}
V_{ud}^2(1+3\lambda^2)=\frac{4908.7(1.9)s}{\tau_n}\end{eqnarray}
see \cite{I}, where $\tau_n$ is the neutron mean life and $\lambda=g_A/g_V$. In our approach
 $V_{ud},\;\Delta_R^V,\;{\rm and}\; \lambda$  are free parameters to be found from fit.

In SM the purely leptonic decay of a $P$ meson, $P\rightarrow l \bar{\nu_l}$, proceeds via annihilation of 
 the quark pair  to a charged lepton and neutrino through exchange of a virtual $W$ boson,
 and the branching fraction, up to radiative corrections,   has the form 
\begin{eqnarray}
{\mathcal{B}}(P\rightarrow  l \bar{\nu_l})=
\frac{G_F^2M_Pm_l^2}{8\pi \hbar}\left(1-\frac{m_l^2}{M_P^2}\right)^2f_P^2 V_{qq'}^2 \tau_P\end{eqnarray}
where $G_F$ is the Fermi constant, $M_P$ and $m_l$  are  the $P$ meson  and  $l$ lepton masses, respectively, 
$f_P$ is the decay constant, $V_{q q'}$ is the modulus of the corresponding KM matrix element, and $\tau_P$ 
is  $P$ lifetime.

 The next simple decays involving $V_{ij}$ moduli are 
 the semileptonic decays of   heavy pseudoscalar mesons, $H$, into  lighter ones, $P$, whose physical observable  
  is the differential decay rate,  written as
\begin{eqnarray}
\frac{d\,\Gamma(H\rightarrow P\,\ell\,\nu_{\ell})}{d q^2}=\frac{G_F^2\,V_{qq'}^2}{192\pi^3 M_H^3}\lambda^{3/2}(q^2)|f_+(q^2)|^2\label{g4}
\end{eqnarray}
where $q=p_H-p_P$ is the transferred momentum,
and $f_+(q^2)$ is  the global form factor which is a combination of  the two   form factors generated by the
 vector part of the weak current.
When the leptons are electrons, or muons whose masses are low compared to the mass difference, $m_H-m_P$,
 $\lambda(q^2)$ is the usual triangle function
\begin{eqnarray}\lambda(q^2)=(M_H^2+M_P^2-q^2)^2-4 M_H^2 M_P^2 \end{eqnarray}

For the decay $\bar{B}\rightarrow Dl\nu$ the experimenters make use of an other variable,
 namely ${\it w}=(M_B^2+M_D^2-q^2)/(2M_BM_D)$. When the $\tau$ lepton is involved  the above formulae are a 
little bit more complicated, see \cite{CLN}.

 Usually, till now, the experimenters provided numerical values for  products of the form $V_{qq'}\,f_+(0)$, and a few for
 $V_{qq'}\,|f_+(q^2)| $.

The second $\chi^2-$component 
which takes into account the experimental data has the form 
\begin{eqnarray}
\chi^2_2=\sum_{i}\left(\frac{d_{i}-\widetilde{d}_{i}}{\sigma_{i}}\right)^2\label{chi2} 
\end{eqnarray}
where  $d_{i}$ are the physical parameters one wants to be found from fit, $\widetilde{d}_{i}$ are  the
 numerical values that describe the corresponding experimental data, while  $\sigma_i$ is the  uncertainty  
associated to $\widetilde{d}_{i}$. For semileptonic decays $d_i$ could be of the form  $d_i= |f_+(q^2_i)|V_{kl}$ .
 In the following our fitting  $\chi^2$-function will be 
\begin{eqnarray}
\chi^2=\chi^2_1 +\chi^2_2\label{chi}\end{eqnarray}

As it is easily seen  the $V_{ij}$ moduli enter naturally in all the  formulae that describe the leptonic and semileptonic
 decays being, in our opinion, a strong argument for their use as fit parameters.

\section{Experimental Data}

In our analysis we used the data on superallowed $0^+  \rightarrow 0^+$ nuclear $\beta$ decays published in the papers \cite{HT1}-\cite{GS}, 
and data on the neutron lifetime  from four papers: \cite{AS}, \cite{JN}, \cite{MD}, \cite{SAr}, for $V_{ud}$ 
determination. We also used  four values for the  $\beta$-asymmetry parameter  $A_0$, from papers \cite{RP},
 \cite{HA}, 
\cite{HA1}, \cite{BY}, and one  for the electron-antineutrino correlation coefficient $a_0$, \cite{JB},
 for the $\lambda$ determination.

$V_{us}$ modulus is involved in the kaon and pion leptonic and semileptonic decays, but also in the ratio
 $V_{us}/V_{ud}$, that appears in the Marciano relation, \cite{WJM}, that we write as
\begin{eqnarray}
\frac{V_{us}^2f_K^2}{V_{ud}^2f_{\pi}^2}(1+C_r)=
\frac{{\mathcal{B}}(K\rightarrow \mu\bar{\nu}_{\mu}(\gamma))\tau_{\pi}m_{\pi}(1-\frac{m_{\mu}^2}{m_{\pi}^2})^2}{{\mathcal{B}}(\pi\rightarrow \mu\bar{\nu}_{\mu}(\gamma))\tau_{K}m_K(1-\frac{m_{\mu}^2}{m_{K}^2})^2}\label{mar}
\end{eqnarray}
where $C_r$ is a radiative correction stemming from both $\pi$ and $K$ hadronic structures. In fact relations of the form (\ref{mar}) can be written for all the leptonic and semileptonic decays, which leads to a cancellation of the experimental errors appearing on the right side. 

 The last  numerical values for the product $f_+^{\pi K}(0)V_{us}$ are given by KLOE collaboration,  
 \cite{Kl}, and by FlaviaNet Working Group, \cite{Fl}, and in fit we made use of all the (little)
 different $f_+(0)V_{us}$ values corresponding to  the five channels. We also used  results from 
\cite{MT}, \cite{JR}, \cite{CB}, \cite{FA}, \cite{Ts}, \cite{Vr}, \cite{Mm}, \cite{FA2}, \cite{FA3},  \cite{MT1}, 
\cite{AB}, \cite{TA} that give only a ``mean value'' for the above product.

 $V_{ub}$ is  the most poorly determined modulus although there is much experimental information coming from 
 decays $B\rightarrow \pi l\nu$, see \cite{SBA}, \cite{TH}, \cite{KI}, \cite{NEA},  \cite{BA2}, \cite{FN}, 
\cite{BA3}. In these papers the experimenters have been confronted with the known difficulty, 
 getting two distinct parameters, $V_{ub}$ and $ |f_+^{B\pi}(q^2)|$, from their product measured from 
experimental data. Thus they used the form factor lattice computations to obtain  $V_{ub}$ values depending 
on $q^2$, see \cite{SBA}, \cite{TH}, and \cite{BA3}. For fit  we found three  measurements for  $f_BV_{ub}$, 
\cite{KI}, \cite{BA1}, \cite{BA2}, and three fenomenological determinations involving
 $f_+^{B\pi}(0)V_{ub}$, \cite{FN}, \cite{BA3}, \cite{BLC}.

$V_{cd}$ and $V_{cs}$ moduli enter the leptonic decays  $D\rightarrow l\nu$, \cite{BE}, \cite{AR}, \cite{ID},
 and $D_s^+\rightarrow l\nu$, \cite{JPA}, \cite{PU}, \cite{KME}, \cite{LW}, \cite{AR1}, respectively, as well 
 as the semileptonic decays  $D\rightarrow \pi l\nu$, and $D\rightarrow K l\nu$, \cite{DCH}, \cite{SD}, \cite{JG}. In the last three papers one find $V_{cq} f_+(0)$
 values, for $q=d$, and  $q=s$,  and for the first time numerical values for the product 
\begin{eqnarray}
|f_+(q^2)|V_{cq}=\sqrt{\frac{d\Gamma}{dq^2}\frac{24 \pi^3 p^3_{K,\pi}}{G_F^2}}\label{ff}
\end{eqnarray}
\cite{JG}, that allow the  form factor extraction    directly from data. The above semileptonic decays also
 allow the measurement of the ratio
$V_{cd}f_+^{D\pi}(0)/V_{cs}f_+^{DK}(0)$, see \cite{SD}, \cite{JG}, \cite{LW1}, \cite{GSH}, \cite{JML}, \cite{MAb},
 and give an independent determination  of the ratio  $f_+^{DK}(0)/f_+^{D\pi}(0) $, which in fit was considered a new 
independent parameter.

Finally from the semileptonic decays $\bar{B}\rightarrow D l\nu$  and  $\bar{B}\rightarrow D^* l\nu$, \cite{BA5},
 \cite{BA6}, \cite{BA7}, \cite{BA}, \cite{IA},  \cite{BA9}, \cite{BA10}, \cite{JAb}, \cite{NEA1}, \cite{KAb}, 
\cite{PA},  \cite{GA}, one find the  $V_{cb}, \;\mathcal{G}(1)$ and  $\mathcal{F}(1)$ parameters.

\section{Numerical results}

Data from  the above cited papers were used to define  $\chi_2^2$, the second component of full $\chi^2$, which 
has a parabolic form in $V_{ij}^2$.  The first component $\chi_1^2$,  (\ref{chi1}), that contain all unitarity
 constraints, has a  parabolic part, and one that is highly nonlinear in all $V_{ij}$. Thus we had to test the
 stability of the 
 expected physical values against the strong non-linearity implied by unitarity.
Eventually the chosen method was to modify all the measured central values in the same sense, plus and minus, 
respectively, proportional to  their corresponding uncertainties. 

An important assumption included in  our approach was: the   numerical $V_{ij}$ moduli values must be the same 
irrespective of the physical processes where they are implied.  Accordingly  the other parameters, such
 as decay constants, $f_P$, form factors,  $f_+(0)$, $\lambda$, etc., that parametrize each given experiment, 
have been considered as independent  parameters to be obtained from  fit, by  applying the usual technique 
to obtain their mean values and uncertainties. 

The stability tests provided sets of different moduli matrices that have been
 used to  obtain a mean value matrix and its corresponding error matrix. The mean and uncertainty matrices 
have been  computed by embedding the  unitary matrices into the double stochastic matrix set, see papers \cite{PD} and \cite{PD1}.

The central values and uncertainties of data used in fit are those published in the above cited papers,
 and we combined the statistical and systematic  uncertainties in quadrature when the experimenters provided 
both of them.
The  numerical values obtained  from fit for  the decay constants, $f_{\pi},\,f_K,\,f_B,\, f_D,\, f_{D_s}$, 
the semileptonic form factors $f_+(0)$,  $\Delta_R^V,\; \lambda$, as well as for the ratios $f_K/f_{\pi} 
{\rm and},$ $\ f_+^{DK}(0)/f_+^{D\pi}(0)$ are given in Table 1. All of them are in the expected range although
 many have big uncertainties.
 
 Our approach allows a  ``fine structure analysis'' of all experiments measuring one definite quantity, such
 as $\Delta_R^V$, or $f_+^{ K\pi}(0)$, etc. 
For example  KLOE collaboration data, \cite{Kl}, and FlaviaNet Working Group data, \cite{Fl}, on $f_+(0)V_{us}$ 
lead  to
\begin{eqnarray}
f_+^{K \pi }(0)_{KLOE} = 950.38 \pm 5.56\;\; {\rm and} \;\;
 f_+^{K \pi }(0)_{Flavia} = 955.06 \pm 4.31\end{eqnarray}
respectively, whose central values are a little bit different, but compatible between them at $1\sigma$,
 and all together provide
\begin{eqnarray}f_+^{K \pi }(0)_{KLOE+Flavia}=952.72 \pm 5.30\end{eqnarray}
$f_+^{K \pi }(0)$ from Table 1 has a precision of 1\%, and the above value obtained from  ten measurements has
 a precision of 0.56\%.
\vskip3mm
Table 1. Numerical values for decay constants $f_P$ and\\
$~~~~~~~~$ form  factors $f_+(0)$ in MeV units, and $\Delta_R^V$ and $\lambda$
 \vskip2mm
\begin{tabular}{|l|l|l|}
\hline
Parameter&Central Value& Uncertainty\\
\hline
$f_{\pi}$ &  131.131 &1.522\\
$f_K$&  154.97 &  2.17\\
$f_K/f_{\pi}$ &  1.1818 &  0.0042\\
$f_B$ &222.8 &25.0\\
$f_D$& 207.6 &9.8 \\
$f_{D_s}$ &271.0 &18.0\\
$f_+^{K\pi}(0)$ &955.34 &9.27\\
$f_+^{B \pi }(0)$ &214.9 &13.4\\
$f_+^{D \pi}(0)$ &653.2 &19.1\\
$f_+^{D K}(0)$ &751.8 &10.4 \\
$f_+^{D K}(0)/f_+^{D \pi}(0)\footnotemark[1]$&1.171&0.049\\
$\mathcal{F}$(1)& 957.5 &57.7\\
$\mathcal{G}$(1) &1,125.3 &40.7\\
$\Delta_R^V\%$ &2.373 &0.096\\
$\lambda$ &-1.2686 &0.0057\\
$C_r$ &$0.002$ &$0.0001$\\
\hline
\end{tabular}
\vskip2mm
\footnotemark[1] {Data on $V_{cd} f_+^{D\pi}/V_{cs} f_+^{D K} $ from  papers  \cite{SD}-\cite{MAb}}

\vskip4mm

In this approach one could obtain  information about lepton universality in leptonic decays. Because the
 corresponding $f_M$ decay constant should be same for both decays $M\rightarrow \mu \nu$ and 
 $M\rightarrow \tau \nu$, a big difference between them could show a possible violation. As an example we 
chose $M = D_s^+$ meson since  lattice computations provided a number for this decay constant,  
 $f_{D_s}=241(3)$, see paper \cite{EF}, with a   very small error. Our results are 
\begin{eqnarray}
f_{D_s^+\rightarrow  \mu \nu} = 265.0\pm 14.0,\;\;
f_{D_s^+\rightarrow  \tau \nu} = 276.0\pm 20.0\end{eqnarray}
which are consistent with lepton universality. The above two numbers together with that from Table 1 completely 
disagree with that provided by lattice computations, being far away from theoretical prediction at 
$8\sigma$,  $13\sigma$,  and $10\sigma$, respectively, 
 where, $\sigma=3$, is the lattice uncertainty. The experimental spreading is, $246.0 \le  f_{D_s^+}\le 311.0$, 
and the minimal and maximal values correspond to the branching ratios obtained for 
 $\mathcal{B}(D_s^+\rightarrow \mu^+ \nu_{\mu})=(5.15 \pm 0.63 \pm 0.20 \pm 1.29)\times 10^{-3}$,
 see \cite{BA4}, and to the branching ratio  $\mathcal{B}(D_s^+\rightarrow \tau\nu)=(8.0 \pm 1.3 \pm 0.6)\%$, given by Eq.(6) in paper \cite{AR1}, respectively. The new result obtained in \cite{CTD} does not improve the situation.

 On the other hand if one computes the difference between 
each one of the above three values and that provided by lattice computation, divided by the corresponding 
experimental error $\sigma$  obtained from  the fit one finds the same value, $1.17$, which shows again the 
consistency of  experimental data. In our opinion the lattice number is much  underestimated. In contradistinction
 their value for $f_K/f_{\pi} = 1.189(7)$ is not far from the fit value $f_K/f_{\pi} = 1.1818(42)$.

Another unexpected  result concerns the $\Delta_R^V$ constancy, usually assumed in all  the four papers
 \cite{HT1}-\cite{GS}, assumption that is not  confirmed by our analysis. However our result
 $\Delta_R^V=2.373 \pm 0.096$ obtained from data \cite{HT1} is in   good concordance with  the value given by 
relation  (\ref{rad}), see Table 2, but   our uncertainty    is $2.5 \sigma$ higher  than the theoretical one.
 In this case the $\Delta_R^V$  spreading that results from paper \cite{HT1} is,  
$2.2027 \le\Delta_R^V\le 2.472$, which corresponds to $7.1 \sigma$  where $\sigma=0.038\%$ is the theoretical
 uncertainty. The extremal nuclei are   $^{22}$Mg and $^{54}$Co, respectively. The spreading obtained
 from Savard \textit{et al} data, \cite{GS}, corresponds to  $10.8 \sigma$, and the extremal nuclei
 are $^{74}$Rb and $^{34}$Cl. However the difference between the central results from \cite{HT1} and 
\cite{GS} is 2$\sigma$, which suggests that there is still room for numerical improvements.  
The  mean values and the corresponding uncertainties for all data from papers \cite{HT1}-\cite{GS} 
are given in Table 2.

\vskip3mm
Table 2.  $\Delta_R^V$ central values and uncertainties provided by fit when using 
$~~~~~~~~~~~~~~~~~$ data from Refs. \cite{HT1}-\cite{GS}
\vskip2mm
\begin{tabular}{|cccccc|}\hline
Ref.&\cite{HT1}&\cite{HT2}&\cite{HT3}&\cite{GS}&\cite{HT2}\\
\hline
$\Delta_R^V\%$&2.373(96)&2.399(112)&3.361(196)&2.294(136)&2.361(38)\footnotemark[1]\\
\hline
\end{tabular}
\vskip2mm
\footnotemark[1] {The ``constant'' $\Delta_R^V$ and its uncertainty, from \cite{HT2}}
\vskip4mm

The precision of $f_P$ and $f_+(0)$ determinations in Table 1 varies from 1\% for $f_+^{K\pi }(0)$ to 11\% for $f_B$. 
More about variability of the above parameters could be  learnt  from KM moduli matrix. Our fit result for 
 KM  central moduli values is 
\begin{eqnarray}
V_c=\left[\begin{array}{lll}
0.974022&0.226415&0.004251\\
0.226253&0.973323&0.038108\\
0.009569&0.037131&0.999265\end{array}\right]\label{centr}
\end{eqnarray}

 $V_c$ and  its associated uncertainty matrix, $\sigma{_{V_c}}$, have been obtained by using the convexity property, (\ref{con}), and  $\sigma{_{V_c}}$ has the form
\begin{eqnarray}
\sigma{_{V_c}}=\left[\begin{array}{lll}
1.1\times 10^{-6}&1.9\times 10^{-5} &2.1\times 10^{-5} \\
3.6\times 10^{-5} &2.7\times 10^{-4} &3.1 \times 10^{-4}\\
3.5\times 10^{-5} &2.9\times 10^{-4} &3.9\times 10^{-4} \end{array}\right]\label{sig}
\end{eqnarray}

The last matrix has been obtained with the help of stabilty tests. One such  matrix is (\ref{sig1}) that was obtained when all the central measured values have been modified  with plus one tenth from the corresponding uncertainty. Although such a modification is highly improbable from an experimental point of view, it brings to light the variation direction for all parameters entering the fit.

\begin{eqnarray}
V_+=\left[\begin{array}{lll}
0.974021&0.226332&0.007484\\
0.226114&0.973083&0.044512\\
0.012432&0.043391&0.998891\\
 \end{array}\right]\label{sig1}
\end{eqnarray}

For example the above three matrices show that $V_{ud}$ is precisely determined with four digits, while 
 $V_{us},\;V_{cd},$ and $V_{cs}$ only with  three digits.
 $V_+$ matrix also shows the high $V_{ub}$ volatility, such that future data could lead to higher values for
 it than that given by $V_c$ matrix. In fact matrices $V_+$ and $V_c$ show that $f_B \in (123.5, 222.8)$. 

Although
 our value for
$V_{ub}=(4.25 \pm 0.02)\times 10^{-3}$ is higher than that from  PDG fit, \cite{pdg08}, there are experimental determinations which are   compatible with it, one example being   Ref. \cite{BA3}, whose value is $V_{ub}= (4.1 \pm 0.2_{st} \pm 0.2_{syst}\,^{+0.6}_{-04.FF})\times 10^{-3}$.

 The new data from $D$ leptonic, \cite{BE}-\cite{ID}, and semileptonic decays, \cite{DCH}-\cite{JG},  combined 
with the new data from $\bar{B}\rightarrow Dl\nu$  and  $\bar{B}\rightarrow D^*l\nu$ \cite{BA5}-\cite{IA},
 changed the KM moduli values, in particular those from the last row and column, see  \cite{pdg08}, p. 150. 

Our approach allows the form factor determination. The paper \cite{JG} provided for the first time results on 
$V_{cq} |f_+(q^2)|$, $q=d,\,s$, and in Table 3
are given our determinations.

A big step forward will be the  measurement of  $q^2$ dependence for products of the  form $|f_+(q^2)U_{qb}|$,
 where $q=u,\,c$, for $B\rightarrow \pi l \nu$, $\bar{B}\rightarrow Dl\nu$  and  $\bar{B}\rightarrow D^*l\nu$
 decays, similar   to that done for $D$ semileptonic decays.
\vskip4mm

 Table 3. $|f_+(q^2)|$ form  factors  from 
 $D\rightarrow \pi l \nu$ and
 $D\rightarrow K l \nu $  decays 

\begin{eqnarray}\begin{array}{cccc}
\begin{tabular}{|c|rr|}
\hline
Bin& $f_+^{\pi}(q^2)\;(\pi^{0}e^+\nu_e)$&$ f_+^{\pi}(q^2)\;(\pi^-e^+\nu_e)$\\
\hline
1&0.6895 $\pm$ 0.0576&0.7072 $\pm$ 0.0355 \\
2&0.7514  $\pm$ 0.0709&0.7735 $\pm$ 0.0400\\
3&0.8442 $\pm$ 0.0841&0.7956 $\pm$ 0.0488\\
4&0.8928  $\pm$ 0.0974&0.9812 $\pm$ 0.0532\\
5&1.1005  $\pm$ 0.115&1.017  $\pm$ 0.065\\
6&1.3083 $\pm$ 0.151&1.101  $\pm$ 0.084\\
7&1.5779  $\pm$ 0.221&1.635  $\pm$ 0.111\\
8&-&1.759  $\pm$ 0.2358\\
9&-&2.024  $\pm$ 0.301\\
\hline
 &$f_+^{K}(q^2)\;(K^{-}e^+\nu_e)$&$ f_+^{K}(q^2)\;(\bar{K}^0e^+\nu_e)$\\
\hline
1& 0.7798 $\pm$ 0.0136&0.7808 $\pm$ 0.0208\\
2&0.8281 $ \pm$ 0.0146& 0.8096 $\pm$ 0.0239\\
3&0.8435 $ \pm$ 0.0156& 0.8466  $\pm$ 0.0259\\
4&0.9113  $ \pm$ 0.0198&0.8856  $\pm$ 0.0289\\
5&0.9945  $ \pm$ 0.0229&0.9175 $\pm$ 0.0331\\
6&1.007  $ \pm$ 0.027&1.024 $\pm$ 0.039\\
7&1.128  $\pm$ 0.033&1.149  $\pm$ 0.048\\
8&1.212 $\pm$ 0.042 &1.090  $\pm$ 0.062\\
9&1.303 $\pm$ 0.066&1.233  $\pm$ 0.092\\
10&1.561 $\pm$ 0.165&1.476 $\pm$ 0.210\\
\hline
\end{tabular}\end{array}\end{eqnarray}

\vskip3mm

 The simplest case is that of  $|f_+(q^2)U_{ub}|$ 
because there are measured data  which in principle could be transformed in values for the product 
$|f_+(q^2)U_{ub}|$. 

Such measurements will allow a more precise  determination of the moduli $V_{ub}$ and  $V_{cb}$,
and, by consequence, they will provide better values for all KM
 matrix moduli from the the first two rows, 
and, perhaps, the  first hints for  new physics beyond SM, if any.

The $V_c$ matrix, (\ref{sig1}), provides numerical values for $\delta$ and the angles of the standard unitarity triangle, as follows
\begin{eqnarray}\begin{array}{cccc}
\delta &=& (89.96\, \pm\, 0.36)^{\circ},&\alpha=(64.59 \, \pm\, 0.27)^{\circ},\\
\gamma&=&(89.98 \,\pm\, 0.06)^{\circ},&\beta=(25.49 \, \pm\,0.28)^{\circ}\end{array}\end{eqnarray}

 The $\delta$ value could be interpreted as a maximal violation of
  {\em CP} symmetry because 
$\sin\delta \approx 1$. A surprising change is the shape of 
 the standard unitarity triangle that is now  a right triangle, since  $\gamma\approx 90^{\circ}$. 
The Jarlskog invariant is $J=(3.567 \pm 0.007)\times10^{-5}$.

Similar results are obtained from  $V_+$ matrix even if $V_{ub}$ is almost twice bigger than that from $V_c$; 
they are
\begin{eqnarray}\begin{array}{cccc}
\delta &=& (90.0\, \pm\, 0.2)^{\circ},&\alpha=(54.1 \, \pm\, 0.1)^{\circ},\\
\gamma&=&(90.0 \,\pm\, 0.4)^{\circ},&\beta=(36.0 \, \pm\,0.2)^{\circ}\end{array}\end{eqnarray}
The above results show that $\delta$ and $\gamma$ angles are independent of $V_{ub}$ variation,
 while $\alpha$ and $\beta$ are moderately dependent. Thus experimental results on  the product
 $|f_+(q^2)U_{ub}|$ could lead to a better determination of all the angles of the standard unitarity triangle.

The small angles uncertainties show that $V_c$ and $V_+$ matrices are highly compatible with unitarity constraints 
and all the 165 $\cos\delta$ formulae provide very close each other values for all of them.

A comparison with lattice computations from \cite{CA} shows that $f_+^{D\pi}(0)$ and  $f_+^{DK}(0)$, as well as 
their ratio are in good agreement with the exprimental values obtained by us, and our uncertainties are smaller.


 \section{Conclusion}

In this paper we presented a phenomenological tool that allows determination of the KM moduli, semileptonic form
 factors and  decay constants  directly from experimental data. It is based on a new
 implementation of unitarity constraints that makes use of KM matrix moduli as fit parameters. These  constraints 
 are strong enough and give   a consistent picture of nowadays flavour physics, and until now provide  no signals 
for physics beyond the SM.

A feature of our tool is that all the 
measurable parameters are each other strongly correlated, a little modification 
of one of 
them propagates to all the other parameters, property which 
is a consequence of unitarity constraints.

The new data from the $D$ leptonic and semileptonic decays, and those from
$\bar{B}\rightarrow D(D^*)l\nu$ led to a 
significant change of  KM moduli
  and of standard  unitarity triangle shape.
 A crucial step forward would be the measurement of $B\rightarrow \pi l \nu$ form factors  in bins of $0.5$ 
GeV$^2$, similar to that done in \cite{JG} for $D\rightarrow\pi l\nu$ and $D\rightarrow K l\nu$, that will allow a better $V_{ub}$ determination, and a stabilization of moduli values entering the first 
two rows.

As a final conclusion one can say that  in contradistinction to experiments from other physics branches, those from semileptonic flavour physics have to be considered as a single experiment because only using all of them, they could provide numbers for all the physical quantities: form factors, decay constants, $V_{ij}$ moduli, etc.

{\bf Acknowledgements.}  We acknowledge a partial support from ANCS Contract No 15EU/06.04.2009.

\end{doublespace}
\end{document}